\documentclass{PoS}
\usepackage{subfig}
\usepackage{amsmath}
\usepackage{multicol}
\usepackage{caption}
\usepackage{cite}

\title{SHiP: a new facility with a
dedicated detector for studying
$\nu_\tau$ properties and nucleon
structure functions}

\ShortTitle{SHiP and its neutrino detector}

\author{\speaker{Annarita Buonaura}\thanks{On behalf of the SHiP Collaboration.}\\
        Universit\'a di Napoli Federico II e INFN Napoli\\
		        E-mail: \email{annarita.buonaura@cern.ch}}

\abstract{SHIP is a new general purpose fixed target facility, proposed at the CERN SPS accelerator. In five years, $2\times 10^{20}$ protons of 400 GeV/c momentum will be dumped on a Molybdenum target.  A detector downstream of the target will allow a search to made for long-lived particles with masses below O(10) GeV/c$^2$ foreseen in several extensions of the Standard Model. Another dedicated detector will allow the study of active neutrino cross-sections and angular distributions. The neutrino detector consists of an emulsion target, based on the Emulsion Cloud Chamber technology fruitfully employed in the OPERA experiment. The Emulsion Cloud Chamber will be placed in a magnetic field, with the so-called Compact Emulsion Spectrometer, a few cm thick chamber for the charge and momentum measurement of hadrons. This will provide the leptonic number measurement also in the hadronic tau decay channels. The detector will be hybrid, using nuclear emulsions and electronic detectors for the time stamp of the events and the measurement of the muon momentum. The muon system will also be based on the design of the one used in the OPERA experiment.}

\FullConference{XXIV International Workshop on Deep-Inelastic Scattering and Related Subjects\\
		11-15 April, 2016\\
		DESY Hamburg, Germany}

\begin{document}

\section{Introduction}

The discovery of the Higgs Boson by the ATLAS and CMS collaborations (\cite{higgsAtlas},\cite{higgsCms}) in 2012 was the final validation of the accuracy of the Standard Model (SM) in describing the world of particle physics.
However, the existence of Dark Matter (DM), the problem of the baryon asymmetry and the non zero mass of neutrinos, not explained by the SM, suggest the presence of new physics Beyond the Standard Model (BSM).

While the LHC and other collider experiments can search for new physics at always increasing energies, the SHiP experiment (Search for Hidden Particles, \cite{shipTP}, \cite{shipPP}) can offer a complementary physics program focusing on very weak couplings. 
Indeed, SHiP will look for very weakly interacting non-SM particles with masses in the GeV region, collectively belonging to the so called ``Hidden Sector'' (HS), such as: light sgoldstinos that appear in the breaking of symmetry of the SUSY theory, sterile neutrinos called Heavy Neutral Leptons (HNL) also foreseen in a minimal extension ($\nu$MSM \cite{numusm}) of the SM as heavy (order of GeV) right handed partners of the SM neutrinos and other singlets with respect to the SM.

Thanks to the peculiar properties of the facility, the SHiP experiment can also be considered a SM neutrino factory, in particular of tau neutrinos. Hence, it will host a neutrino detector to discover the tau anti-neutrino and to study tau neutrino and anti-neutrino cross-sections.

\section{The detector to explore the Hidden Sector}

\begin{figure}[h]
	\centering
	\includegraphics[width=0.7\textwidth]{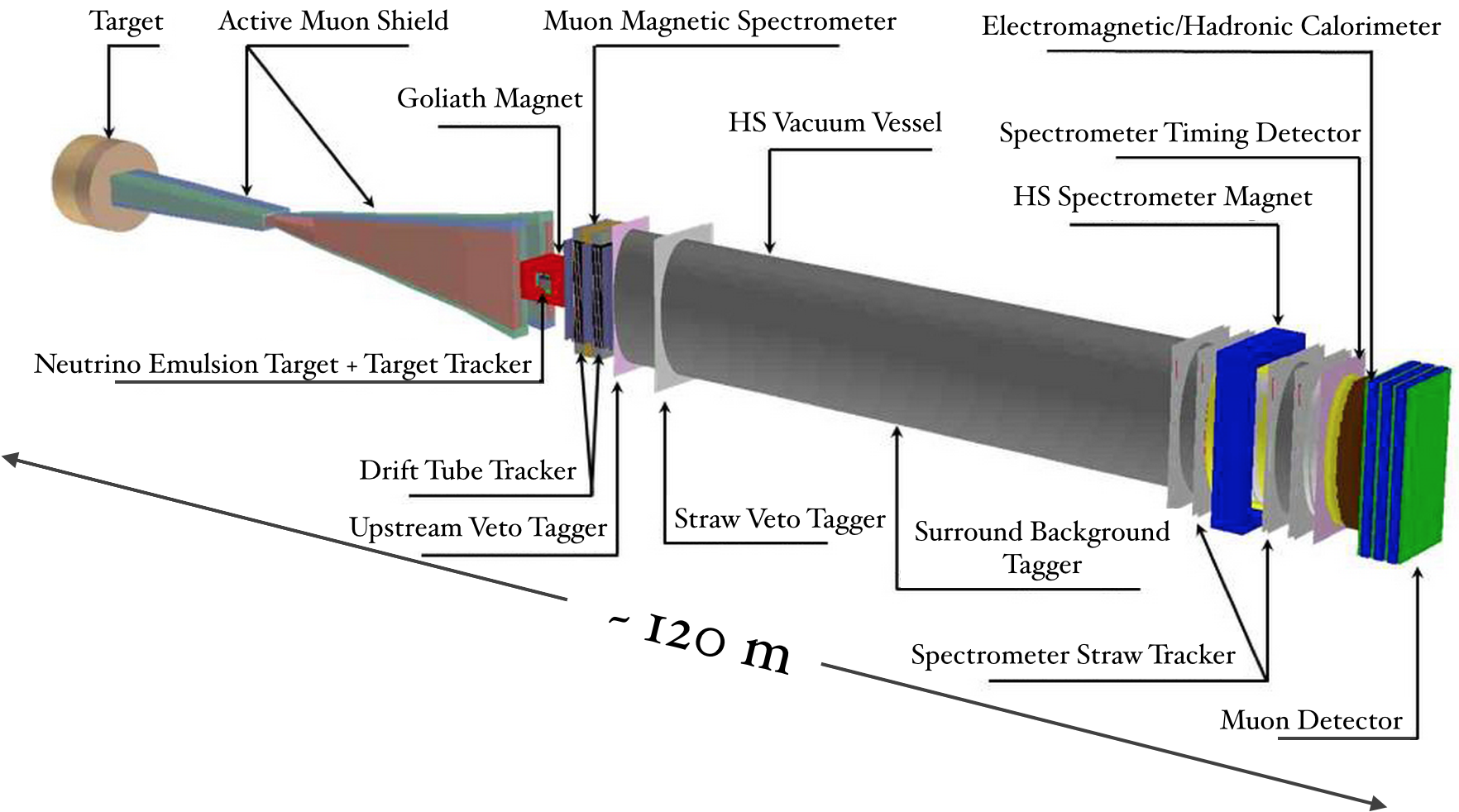}  
	\caption{Overall detector layout of the SHiP facility.}
	\label{fig:Apparatus}
\end{figure} 

In fig.\ref{fig:Apparatus}, the overall layout of the detector is shown. 

400 GeV/c proton spills (4$\times$10$^{13}$ p.o.t.) from a dedicated beam line at the SPS accelerator will impinge on a hybrid target made of blocks of titanium-zirconium doped molybdenum (TZM) followed by blocks of	pure tungsten, materials with a short interaction length, to maximize neutrinos from charmed hadrons, to which Hidden Particles (HP) couple, while minimizing those coming from pions and kaons.

\begin{figure}[h]
	\centering
	\subfloat [\emph{Example of a charmed hadron decay producing Heavy Neutral Lepton}.]{\includegraphics[width=0.45\columnwidth]{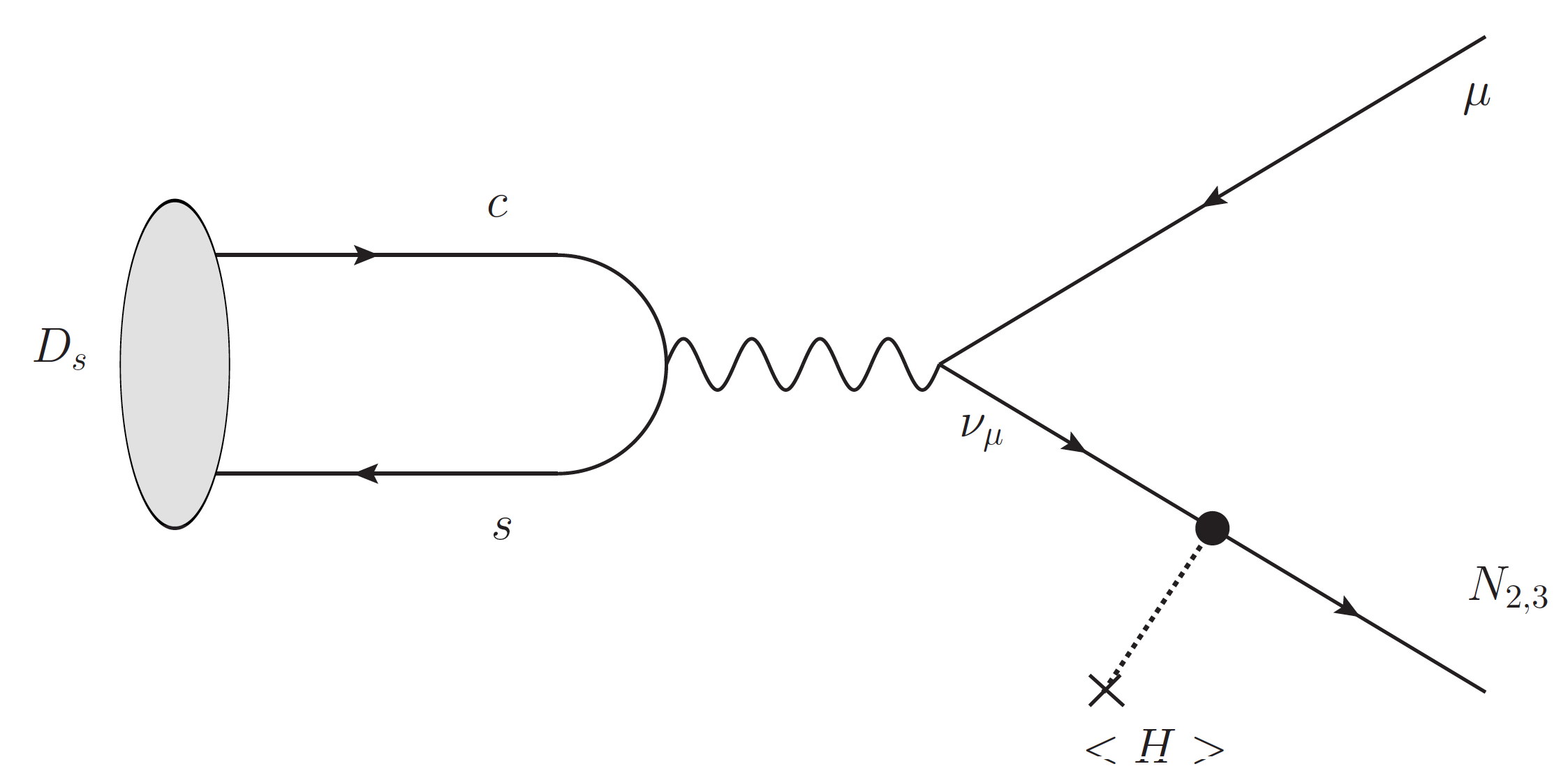}} \quad
	\subfloat [\emph{Total $c\bar{c}$ production cross sections at fixed target energies} \cite{lourenco}, \cite{abramowicz}, \cite{olive}.]{\includegraphics[width=0.35\columnwidth]{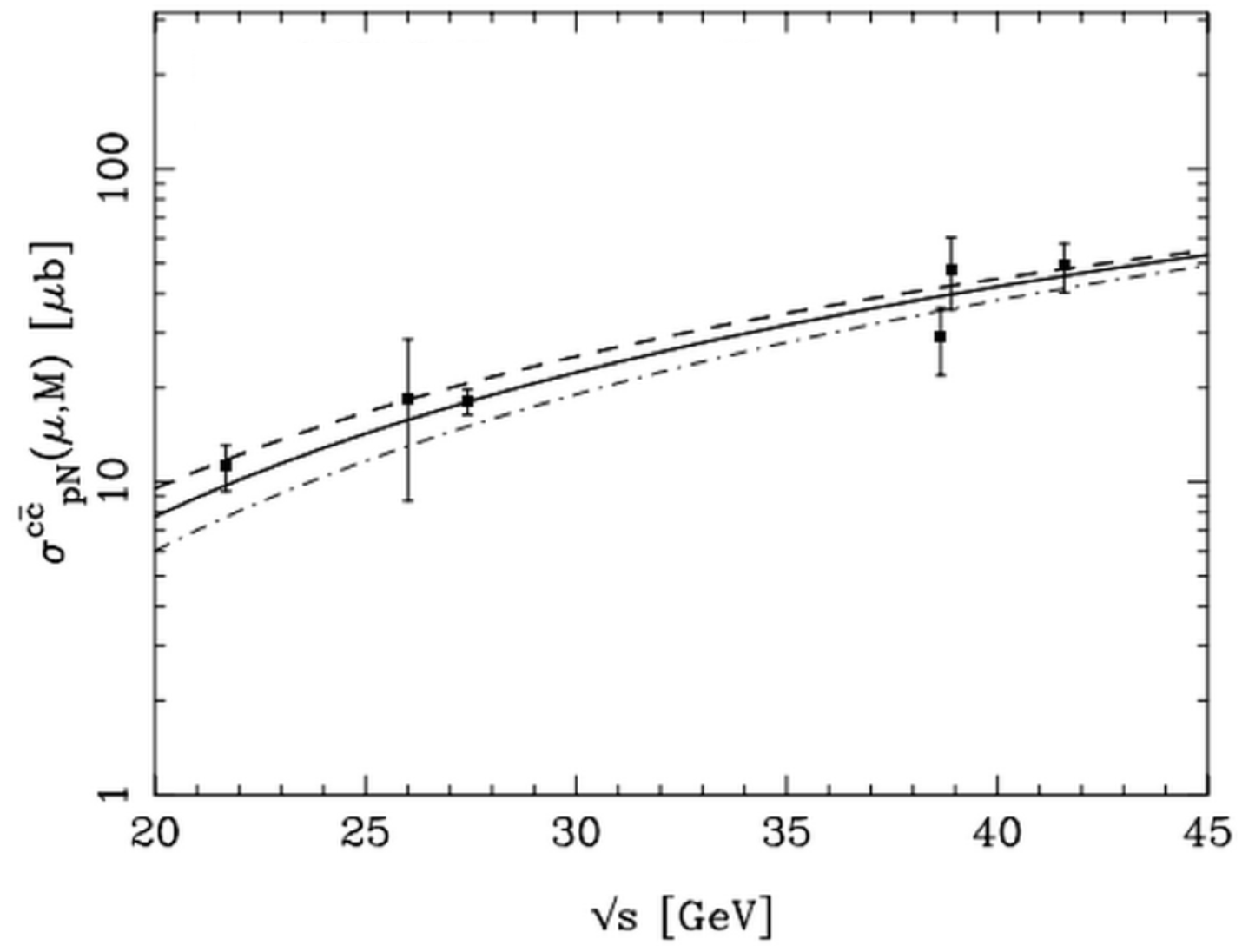}} 
\caption{Example of a production mechanism of HNLs and $c\bar{c}$ cross-section for different $\sqrt{s}$}
\label{fig:prodAndcs}
\end{figure}

The choice of a proton beam with an energy of 400 GeV ($\sqrt{s} = 27 GeV$) is the optimal solution because at this energy the production cross sections of $c\bar{c}$ tend to saturate (fig.\ref{fig:prodAndcs}b) \cite{lourenco, abramowicz, olive}. 

Also the length of the target is chosen to contain possible hadronic showers with minimum leakage and, to have a greater efficiency, a hadron stopper made of 5m of Iron is placed downstream.

HP are then supposed to decay in SM particles (tab.~\ref{tab:decayHP} reports their different decay modes according to the theoretical model considered) with typical lifetimes larger than 10 $\mu$s and corresponding decay distances of the order of the km. 

\begin{table}[h]
	\centering
	\begin{tabular}{cc}
		\hline
		Models & Final states\\
		\hline
		Neutrino portal, SUSY neutralino & $l^{\pm}\pi^{\mp}$, $l^{\pm}K^{\mp}$, $l^{\pm}\rho^{\mp}$, $\rho^{\pm}\rightarrow \pi^{\pm}\pi^{0}$\\
		Vector, scalar, axion portals, SUSY sgoldstino & $l^{+}l^{-}$ \\
		Vector, scalar, axion portals, SUSY sgoldstino & $\pi^{+}\pi^{-}$, $K^+ K^-$ \\
		Neutrino portal, SUSY neutralino, axino & $l^{+}l^{-}\nu$\\
		Axion portal, SUSY sgoldstino & $\gamma \gamma$\\
		SUSY sgoldstino & $\pi^0 \pi^0$\\
		\hline
	\end{tabular}
	\caption{Summary of the main decay modes of hidden particles in various models ($l = e,\mu$)}
	\label{tab:decayHP}	
\end{table}

Therefore, to detect the decay products of these very long-lived particles, a 50 m long decay vessel, placed $\approx$60 m far from the beam dump and equipped with a magnetic spectrometer, calorimeters and muon detectors at the far end is foreseen.
However, since HP produced in the decay of charmed hadrons show a significant transverse momentum, to maximize the acceptance it is important to place the decay volume as close as possible to the proton target. Optimizations are under study in preparation for a future Technical Design Report.

The main source of background is the muon flux coming from short-lived resonances and from the decay of residual pions and kaons. The reduction of this type of background is also of great advantage for the neutrino physics program where precise limits on the acceptable muon rates are set from the use of a nuclear emulsion target. 
To clear a 5 m horizontally wide region, a 48 m long active muon shield based on magnetic deflection of the muons in the horizontal plane is introduced right after the hadron stopper.

Other backgrounds are the combinatorial background from the residual muons entering the decay volume, muons deflecting off the cavern walls and SM neutrinos which can interact in the material upstream of the fiducial volume (the vessel is kept at low pressure ($10^{-6}$ -- $10^{-5} bar$) with subsequent production of charged and neutral particles.
 To prevent them, the proton spills are prepared with a slow (1s) and very uniform beam extraction and a combination of light taggers is located upstream and at the beginning of the fiducial volume.

\section{The neutrino detector}

A 400 Gev beam is optimized to enhance the production of charmed mesons and in particular of $D_s$ mesons. A high flux of $\nu_\tau$ and $\bar{\nu}_\tau$, together with neutrinos of all flavours, is also expected.

The tau neutrino is the least well understood known particle in the SM.

Discovered in 2000 by the DONUT experiment which observed 9 candidate events out of a background of 1.5 \cite{donut}, so far only 5 other $\nu_\tau$ events have been detected by the OPERA experiment \cite{opera, 1tau, 2tau , 3tau, 4tau} leading to the discovery of $\nu_{\mu}\rightarrow \nu_{\tau}$ oscillations in appearance mode with a significance larger than 5$\sigma$ \cite{5tau}. 

Still, the $\bar{\nu}_\tau$ has never been detected, making it the last missing tile of the SM.

Hence, the neutrino detector (fig.\ref{fig:nudetector}a) of the SHiP experiment is designed to perform the first direct observation of the $\bar{\nu}_\tau$ and, thanks to the expected unprecedented statistics of tau neutrinos and anti-neutrinos, also to study their properties and their cross-section.

\subsection{Experimental requirements and the apparatus} 
\label{sec:nudet_description}

\begin{figure}[h]
	\centering
	\subfloat[ \emph{Layout of the neutrino detector}.]{\includegraphics[width=0.57\columnwidth]{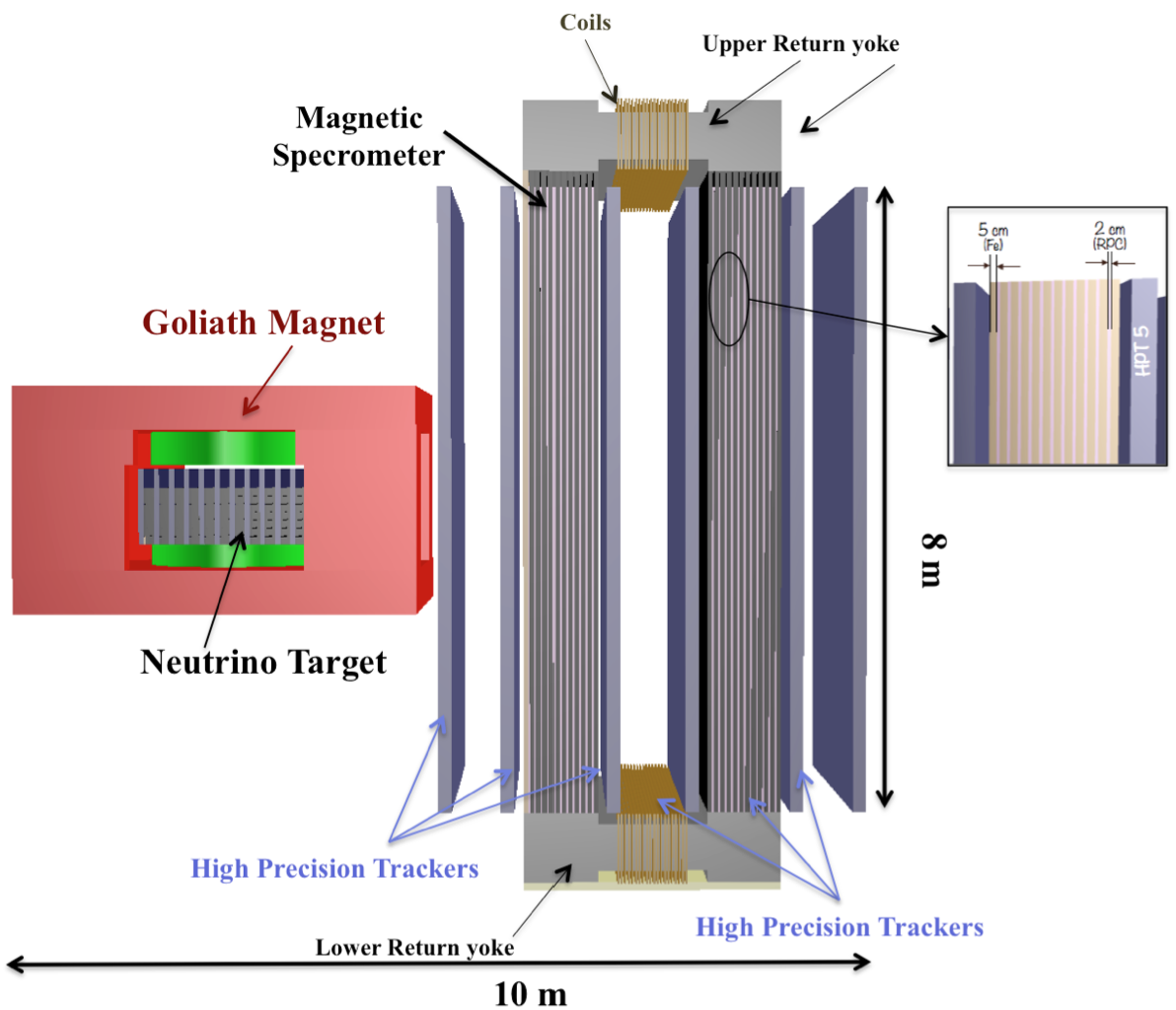}} \quad
	\subfloat[ \emph{Schematic representation of the neutrino detector unitary cell}.]{\includegraphics[width=0.4\columnwidth]{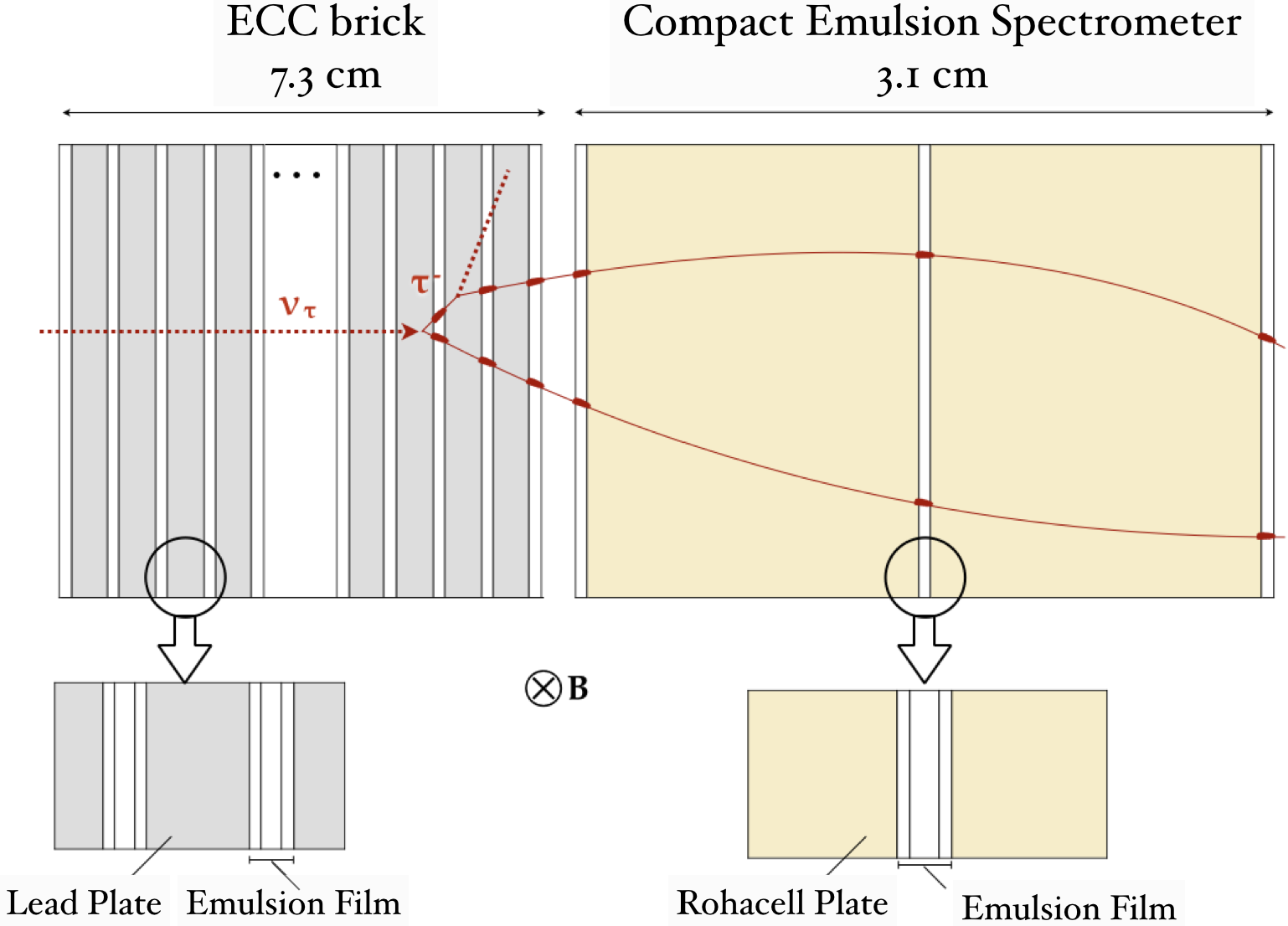}}
\caption{View of the tau neutrino detector and of its fundamental unit.}
\label{fig:nudetector}
\end{figure}

The idea is to locate the neutrino detector in between the active muon shield and the decay vessel.
A neutrino target, placed in a magnet providing a vertical magnetic field ranging from 1 to 1.5 T, is followed by a muon magnetic spectrometer downstream (fig.\ref{fig:nudetector}a).
The neutrino target has a mass of $\sim$9.6 tons and a modular structure. The unit is a ``brick'' with a Compact Emulsion Spectrometer (CES) as shown in fig.\ref{fig:nudetector}b.

\begin{figure}[h]
\centering
\includegraphics[width=0.5\textwidth]{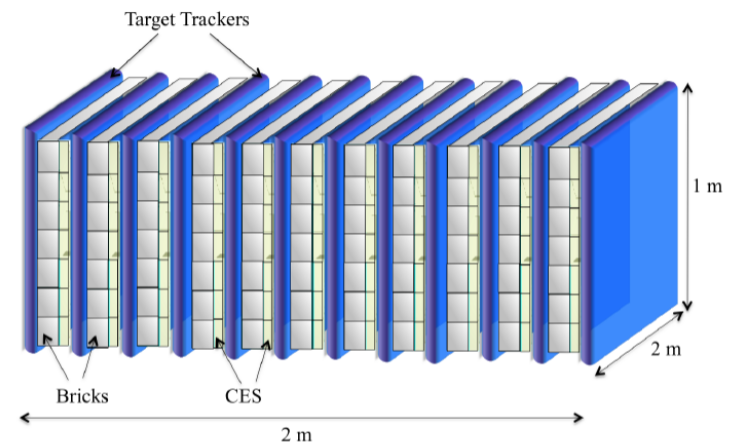} 
\caption{Schematic view of the target with the 12 target tracker layers.}
\label{fig:nutarget}
\end{figure}

The brick (128$\times$ 102 $\times$ 79 mm$^3$, 8.3 kg) employs the Emulsion Cloud Chamber (ECC) technology, largely exploited by the OPERA experiment, interleaving 1 mm lead plates acting as passive material and 300 $\mu$m emulsion films acting as tracking devices with micrometric resolution, which is necessary to disentangle the $\tau$ lepton production and decay vertices.  The bricks are arranged in 11 vertical walls, each made of 105 of these units, as shown in fig.\ref{fig:nutarget}.

In order to detect the charge of the hadronic $\tau$ daughters which do not reach the downstream muon magnetic spectrometer, the CES is attached downstream of each ECC brick. It is made of three emulsion films interleaved with some light material (e.g. Rohacell) 1.5-mm thick and it is designed to perform charge measurement for hadrons with momentum up to 10--12 GeV through the sagitta method.

Each wall of bricks is interleaved with electronic detectors, needed to provide the time stamp of the neutrino interaction occurred in the brick and to link the muon tracks in the target with the magnetic spectrometer. 
Up to now different options are considered as electronic trackers: scintillating fibres or micro-pattern gas detectors like GEM or micromegas. 
These tracking chambers must satisfy very stringent requirements (100 $\mu$m position resolution on both coordinates and high efficiency for angles up to 1 rad) which come from the need of disentangling tracks from the same interaction but coming from two separate vertices and of vetoing penetrating background muons.\\

The muon magnetic spectrometer has the same structure of the OPERA magnetic spectrometer. It is made of 2 arms with opposite 1.5 T magnetic fields. Each arm is made of 12 iron layers, 5 cm thick, interleaved with 11 RPC planes, 2 cm thick, and it is also equipped with 6 drift tube tracker planes. 
It plays a crucial role in identifying muons produced in neutrino interactions and in $\tau$ decays and measuring their charge and momentum. 

Indeed, muon identification at the neutrino interaction vertex is fundamental for background rejection since the main background to the $\tau$ lepton identification comes from the decay of charmed hadrons produced in $\nu_\mu$ charged current (CC) interactions, which exhibit the same decay topology when the primary lepton is not identified.

The opposite magnetic field direction in the two arms ensures a large acceptance, an efficiency of more than 99$\%$ in charge identification and a momentum resolution better than 25$\%$.

\subsection{Physics with the neutrino detector}

\begin{figure}
\centering
\begin{minipage}[b]{.4\textwidth}
\centering
\begin{scriptsize}

\begin{tabular}{c | c c | c c  }
\hline
& $<$E$>$  & Beam dump  & $<$E$>$ &  CC DIS\\
 \hline
  $N_{\nu_\mu}$ & 1.4 & $4.4 \times 10^{18}$   & 29 & $1.7 \times 10^{6}$ \\
   $N_{\nu_e}$     & 3 & $2.1 \times 10^{17}$   & 46 & $2.5 \times 10^{5}$ \\
    $N_{\nu_\tau}$ & 9 & $3.1 \times 10^{15}$    & 59 & $7.4 \times 10^{3}$ \\
  $N_{\overline{\nu}_\mu}$  & 1.5 & $2.8 \times 10^{18}$   & 28 & $6.7 \times 10^{5}$ \\
   $N_{\overline{\nu}_e}$       & 4 &$1.6  \times 10^{17}$   & 46 &$9.0 \times 10^{4}$ \\
    $N_{\overline{\nu}_\tau}$   & 8 & $3.1 \times 10^{15}$  & 58 & $3.7 \times 10^{3}$ \\
 \hline
 \end{tabular}
 \end{scriptsize}
 \centering
 \captionof{table}{Integrated neutrino yield for $2\times10^{20}$ p.o.t. for the  different neutrino flavors: at the beam dump (left) and CC DIS interactions (right). Energies are in GeV.}
 \label{tab:neutrino_all}
 \end{minipage}\hfill
 \begin{minipage}[b]{.57\textwidth}
 \centering
 \subfloat[ \emph{Beam dump}.]{\includegraphics[width=0.47\columnwidth]{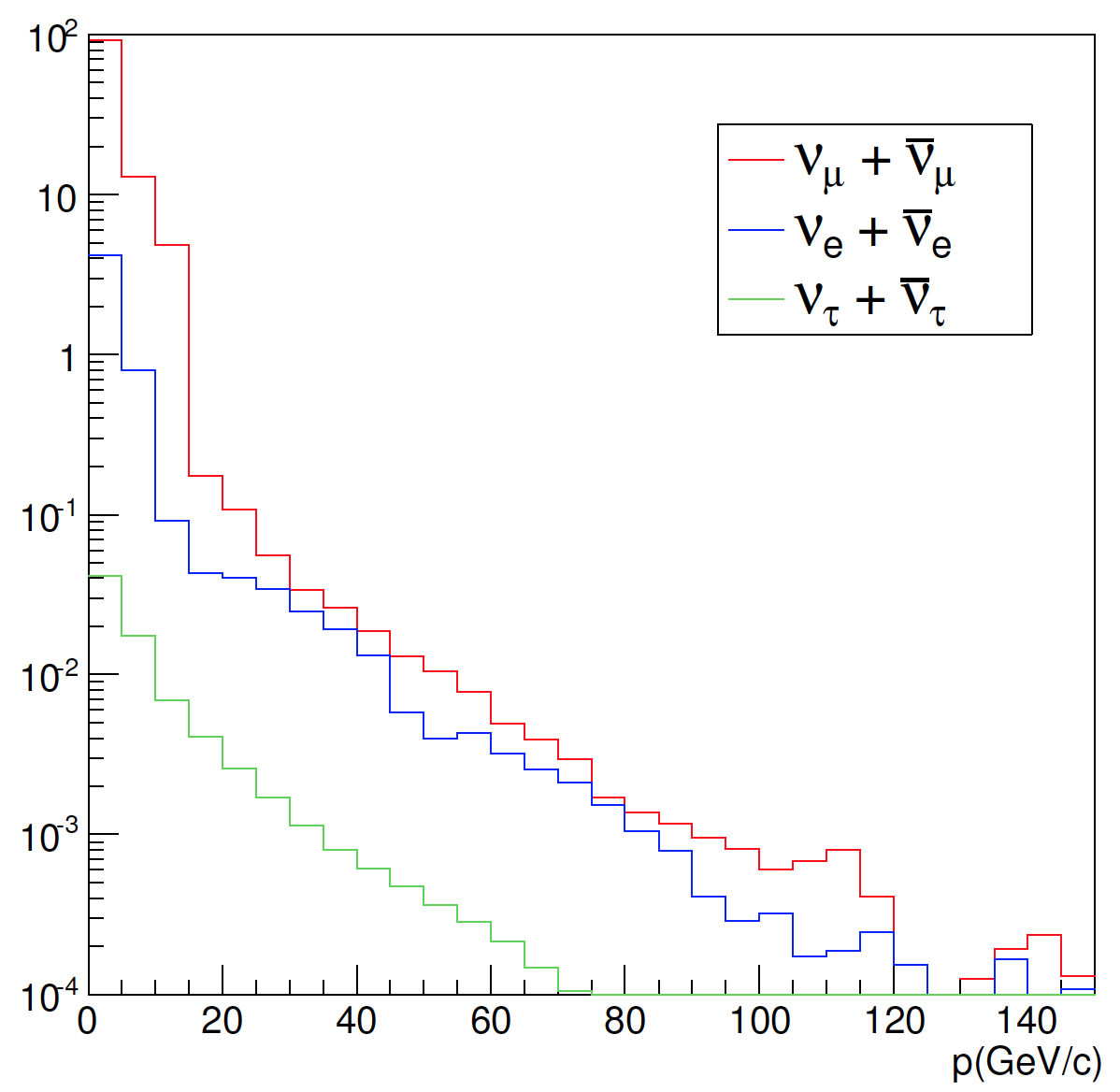}} \quad
 \subfloat[ \emph{CC interacting}.]{\includegraphics[width=0.47\columnwidth]{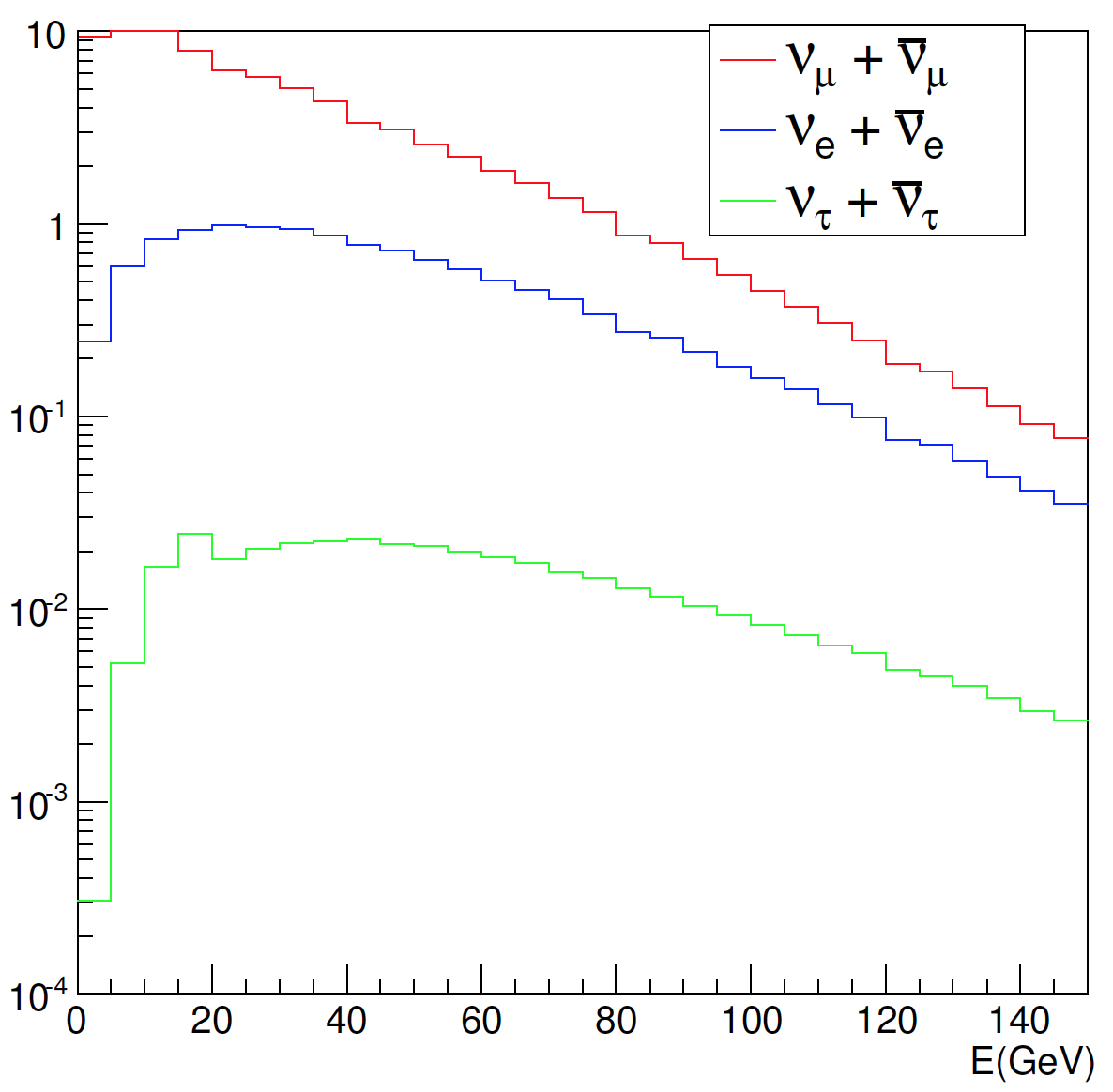}}
 \caption{Energy spectra of the three neutrino flavors at the beam dump (a) and of CC interactions in the neutrino target(c). The total number of neutrinos is normalized to 100.}
 \label{fig:spectra}
 \end{minipage}
 \end{figure}

Tau neutrinos are copiously produced via $D_{s} \rightarrow \tau \nu_{\tau}$ and subsequent $\tau \rightarrow \nu_{\tau}$ decay. In five years running, $6.2 \times 10^{15}$ $\nu_\tau + \bar{\nu}_\tau$ per 2$\times 10^{20}$ protons on target are expected. A high rate of $\nu_e$ and $\nu_{\mu}$, induced by the decay of soft pions and kaons, is also foreseen. The spectra and the neutrino yield at the beam dump are reported in fig.\ref{fig:spectra}a and in the left column of tab.\ref{tab:neutrino_all}, respectively. The expected  number of CC deep-inelastic neutrino interactions during the whole data taking is reported in the right column of Tab.\ref{tab:neutrino_all} while the energy spectra are shown in fig.\ref{fig:spectra}c. All the results have been obtained using a Geant4 based simulation of the facility. It is important to note that almost 8000 $\nu_\tau$ and 4000 $\bar{\nu}_\tau$ charged current interactions are expected, a statistics sufficient to perform cross-section studies. 

From the study of the deep-inelastic scattering (DIS) cross-section of $\nu_\tau$ on nucleons, it is possible to extract the $F_4$ and $F_5$ structure functions, which are negligible in muon (and electron) neutrino interactions because of a suppression factor proportional to the square of the charged lepton mass. At leading order, in the limit of massless quarks and target hadrons, $F_4=0$ and $2xF_5=F_2$, where $x$ is the Bjorken-$x$ variable \cite{aj}. Calculations at NLO show that the $F_4$ contribution to this cross-section is about 1$\%$ \cite{reno}.

Thanks to the large flux of $\nu_e$ and $\nu_\mu$ reaching and interacting in the neutrino target, not only studies of $\nu_\tau$ physics can be performed. The strange-quark content of the nucleon can be measured by means of the charmed hadron production in anti-neutrino interactions. In 5 years running, more than $10^5$ neutrino induced charmed hadrons are expected. The micrometric resolution of the detector allows identification of these hadrons through the decay topology. Not only di-lepton events have thus to be taken into account making the available statistics exceed by more than one order of magnitude that of previous experiments. A simulated SHiP data sample normalized to the expected statistics was used to estimate the distribution of $\nu_\mu (\bar{\nu}_\mu)$ interactions with charm production. The potential impact of simulated charm data was assessed by adding them to the NNPDF3.0 NNLO fit \cite{nnpdf}. Defining $s^{\pm}$ as $s^{\pm} = s(x)\pm \bar{s}(x)$, almost a factor two gain on the accuracy is achieved in the $x$ range between 0.03 and 0.35 for $s^+$ and  between 0.08 and 0.30 for $s^-$.

Experiments performing direct dark matter (DM) searches are still not sensitive to DM particles with masses of O(GeV) or lighter. This mass region can be explored by accelerator experiments, where the target recoils gain enough energy to be detected. In a proton dump, light DM particles ($\chi$) can be produced via the decay of the dark photon (\cite{dm1, dm2, dm3, dm4}) and the neutrino detector is well suited to identify them through the scattering off electrons. Background sources for this search are neutral current $\nu_\mu$ and $\nu_e$ scattering on electrons, and charged current elastic, resonant and deep inelastic $\nu_e$ scattering off nuclei. The main variables to
separate signal from background are the electron energy and the angle with respect to the neutrino direction and the number of detectable particles at the neutrino interaction vertex. Preliminary studies have shown that the sensitivity that the SHiP neutrino detector can achieve considerably extends the reach of previous experiments.

\end{document}